\documentclass[preprint,groupedaddress,floatfix,%
nofootinbib]{revtex4}

\usepackage{euscript}
\usepackage{dcolumn}% Align table columns on decimal point
\usepackage{graphicx}
\usepackage{float}

\usepackage{hyperref}
\usepackage{dcolumn}% Align table columns on decimal point
\usepackage{longtable}
\usepackage{times}
\usepackage{amsmath,amssymb,bm}
\usepackage[english]{babel}
\usepackage[latin1]{inputenc}
\usepackage{times}
\usepackage[T1]{fontenc}
\usepackage{color}
\usepackage{fancybox}
\usepackage{sidecap}
\usepackage{epsfig}
\usepackage{psfrag}
\usepackage{bbm}
\usepackage[english]{babel}
\usepackage[latin1]{inputenc}
\usepackage{times}
\usepackage[T1]{fontenc}
\usepackage{color}
\usepackage{fancybox}
\usepackage{sidecap}
\usepackage{psfrag}
\usepackage{bbm}
\usepackage{wasysym}
\usepackage{slashed}
\usepackage{tikz}
\usepackage{euscript}

\usepackage{algorithm}
\usepackage{algpseudocode}

\begin{document}

\title{Stabilizer Approximation III: Maximum Cut}
\author{Chuixiong Wu}
\author{Jianan Wang}
\author{Fen Zuo\footnote{Email: \textsf{zuofen@miqroera.com}}}
\affiliation{Shanghai MiQro Era Digital Technology Ltd, Shanghai, China}

\begin{abstract}

We apply the stabilizer formalism to the Maximum Cut problem, and obtain a new greedy construction heuristic. It turns out to be an elegant synthesis of the edge-contraction and differencing edge-contraction approaches. Utilizing the relation between the Maximum Cut problem and the Ising model, the approximation ratio of the heuristic is easily found to be at least $1/2$.
Moreover, numerical results show that the heuristic has very nice performance for graphs with about 100 vertices.

\end{abstract}
 \maketitle

\tableofcontents

\section{Introduction}
The Maximum Cut problem for weighted undirected graphs, Max-Cut, is NP-hard. On the contrary, the Minimum Cut problem, Min-Cut, can be solved in polynomial time when the weights are non-negative. Typical algorithms for solving Min-Cut utilize the vertex-identifying and edge-contraction operations~\cite{Cook-1998}. So one may wonder, can we use these operations to construct approximate algorithms for Max-Cut?

In 2007 a greedy heuristic for Max-Cut was conceived using Edge Contraction~(EC)~\cite{Kahruman-2007}. The approximation ratio was prove to be at least $1/3$. However, numerical results show that the heuristic performs worse than the very first Sahni-Gonzalez~(SG) algorithm~\cite{SG-1976} and its improved versions~\cite{Kahruman-2007}.

A variation of the EC heuristic, Differencing Edge-Contraction~(DEC), was proposed recently in~\cite{Hassin-2021}. The relation between EC and DEC is clear: while EC is a ``worst-out'' approach, DEC is a ``best-in'' one. Moreover, even if the original weights are non-negative, negative weights could appear during DEC. This makes it difficult to derive a theoretical approximation ratio for the algorithm. However, numerical analysis shows that its performance could even compete with perhaps the best version of the Sahni-Gonzalez algorithm, SG3~\cite{Kahruman-2007}.

Could we make a synthesis of these two heuristics and obtain a ``best-in-worst-out'' approach? Yes! In the following sections we present the details of such an approach. But before that, we would like to briefly explain how the idea comes out. It has long been known that Max-Cut is closely related to the classical Ising model in statistical physics. In Ising model one tries to obtain the spin state with the lowest energy, the ground state. The spin states could be specified by the spins at the vertices, or by their relations on the edges. Roughly speaking, independent operators generating these relations are defined as ``stabilizer generators'', or simply ``stabilizers''. For an introduction to the stabilizer formalism, see ref.~\cite{Gottesman98}, Chapter 10 of~\cite{Nielsen-Chuang-2010} and references therein. Our heuristic thus proceeds by searching these stabilizers greedily to minimize the total energy of the Ising model. After translating the whole process back into Max-Cut, we forget all these physical interpretations and get a pure graphical algorithm, which we still call ``Stabilizer''.

\section{Stabilizer Heuristic}

\subsection{Max-Cut}

%For a undirected graph $G=(V,E)$ with weights $w: E(G)\to \mathbb{R}$, the partition of $V$ into $V=S\uplus T$ defines a cut %\begin{equation}
%cut(S,T):= E(S,T),
%\end{equation}
%and the corresponding weight:
%\begin{equation}
%w(S,T):= \sum _{e\in cut(S,T)} w(e).
%\end{equation}
%

Given a weighted undirected graph $G=(V,E,w)$ with $V=\{1, 2,...,n\}$ and weights $w: E\to \mathbb{R}$, an assignment $z:V\to \{-1,+1\}$ defines a cut of $G$
\begin{equation}
c(z):=\{(i,j)\in E|z_iz_j=-1, i,j=1,2,...,n\},
\end{equation}
and the corresponding cut weight
\begin{equation}
w_c(z):= \sum _{(i,j)\in c(z)} w_{i,j}=\frac{1}{2}\sum_{1\le i< j \le n} w_{ij}(1-z_iz_j).\label{eq.MC-Ising}
\end{equation}
The Max-Cut problem aims to find the cut with the maximum weight. Actually the above expression for the cut weight reveals the relation between Max-Cut and the Ising model~\cite{BGJR-1988}.

Define
\begin{equation}
U:=\{i|z_i=+1, i=1,2,...,n\},\quad D:=\{i|z_i=-1, i=1,2,...,n\},
\end{equation}
then a cut of $G$ can also be expressed as
\begin{equation}
c(U,D):= E(U,D):=\{(i,j)\in E|i\in U, j\in D\},
\end{equation}
with the weight
\begin{equation}
w_c(U,D):=\sum_{i\in U, j\in D} w_{ij}.
\end{equation}

One may conceive approximate algorithms by constructing $U$ and $D$ iteratively, such as the Sahni-Gonzalez algorithm~\cite{SG-1976} and its improved versions~\cite{Kahruman-2007}. The original SG algorithm operates in a pre-chosen order, and at each step makes the assignment according to the rule~\cite{SG-1976}:
\begin{equation}
\mbox{if}\quad w(i,U)<w(i,D),~\mbox{then}\quad U\leftarrow  U \cup \{i\},~\mbox{else}\quad D \leftarrow  D \cup \{i\}.\label{eq.SG}
\end{equation}
Here $w(i,U)$ and $w(i,D)$ are defined for $i\in V\setminus(U\cup D)$ as:
\begin{equation}
w(i,U):=\sum_{j\in U} w_{ij}, \quad   w(i,D):=\sum_{j\in D} w_{ij}.
\end{equation}
The improved versions~\cite{Kahruman-2007} differ from the original SG by the construction order. In SG3 one first chooses the vertex
\begin{equation}
i^*=\arg \max_{i\in V\setminus(U\cup D)} |w(i,U)-w(i,D)|
\end{equation}
at each step. Then one makes the update of $U$ and $D$ according to (\ref{eq.SG}) as in SG. Both SG and SG3 have the approximation ratio $1/2$, but SG3 outperforms SG a lot numerically~\cite{Kahruman-2007}.

\subsection{Edge Contraction}
The EC heuristic~\cite{Kahruman-2007} bears remarkable similarity to the vertex-identifying and random contraction algorithms for Min-Cut~\cite{Cook-1998}.
At each step, EC chooses the edge, say $(i,j)$, with the minimum weight, and contracts it. Then vertices $i$ and $j$ are replaced by a new vertex $ij$. Multiple edges appear after this operation. For a vertex $k$ distinct from $i$ and $j$, the resulting multiple edge between $k$ and $ij$ is replaced by a new edge, with the updated weight
\begin{equation}
w_{ij,k}=w_{i,k}+w_{j,k}.
\end{equation}
The algorithm terminates when there is only one edge left, say $(u,d)$. Then all the vertices contracted to give $u$ forms $U$, and those contracted to give $d$ forms $D$. The cut weight is simply $w(u,d)$.

The approximation ratio of EC, taking the total weight as reference, is proved to be at least $1/3$. Numerical analysis verifies that it is strictly smaller than $1/2$. For concrete instances, the heuristic could occasionally outperform SG, but on average it performs worse than SG, and much worse than SG3.

\subsection{Differencing Edge Contraction}

Obviously EC belongs to the ``worst-out'' approaches for Max-Cut. The corresponding ``best-in'' approach gives the DEC algorithm~\cite{Hassin-2021}. Therefore, at east step DEC chooses the edge with the maximum positive weight, and contracts it in a differencing way. Contrary to EC, DEC needs to be done with a direction for the edge. For example, if we merge $j$ to $i$, the new vertex will be labeled by the expression $i-j$, and the weight of the resulting multiple edge is updated as
\begin{equation}
w_{i-j,k}=w_{i,k}-w_{j,k}.
\end{equation}
So the resulting weights depend on the contraction directions. It turns out that if one chooses the direction to maximize the total weight, the performance of the algorithm would be better. The algorithm terminates when there is no positive-weight edge remaining. Then $U$ are formed by those vertices with even numbers of minus signs in the contraction labels, and $D$ by those with odd numbers of minus signs.

It seems DEC does not guarantee a lower bound for the approximation ratio. Numerically, it perform strictly better than SG3 for certain instances, but worse for some others~\cite{Hassin-2021}.

\subsection{Stabilizer}
The stabilizer heuristic could be viewed as a synthesis of EC and DEC, even though it does not come out in this way. So it is a ``best-in-worst-out'' approach. The idea is very simple: at each step one chooses the edge in such a way that the absolute value of its weight is maximized. Then, if the weight is negative, we perform EC; if it is positive, we perform DEC. When doing DEC, the direction will be arbitrarily chosen. The details of the algorithm are summarized in the following pseudo-code.

\begin{algorithm}[H]
\caption{The Stabilizer Heuristic}\label{algo:Stabilizer}
\begin{algorithmic}
\State \textbf{Input:} A complete weighted graph $G=(V,E,w)$ with $V=\{1, 2,...,n\}$ and $w: E\to \mathbb{R}$
\State \textbf{Output:} A cut $V=U\uplus D$ and the cut weight $w_c$
\State $V'\gets V$, $E'\gets E$, $S \gets \varnothing$
\State $w_c \gets \frac{1}{2}\sum_{1\le i< j \le n} w_{ij}$
\While{$E'\neq \varnothing$ and $\max_{e\in E'} |w(e)|\ne0$ }
    \State $(i,j)\gets\arg \max_{e\in E'} |w(e)|$
    \State $w_c\gets w_c+\frac{1}{2}|w_{ij}|$, $z_{ij}\gets -\frac{w_{ij}}{|w_{ij}|}$
    \State $S\gets S\cup \{(i,j)\}$, $E'\gets E'\setminus\{(i,j)\}$, $V'\gets V'\setminus \{i\}$
    \For{$k\in V'\setminus \{j\}$}
        \State $w_{jk}\gets w_{jk}+z_{ij}w_{ik}$
        \State $E'\gets E'\setminus\{(i,k)\}$
    \EndFor
\EndWhile
\State $U\gets \varnothing$, $D\gets \varnothing$
\While {$V\setminus(U\cup D)\ne \varnothing$}
    \State select $i\in V\setminus(U\cup D)$, $z_i\gets+1$, $U\gets U\cup \{i\}$
    \State $R\gets\{i\}$, $Q\gets\{i\}$
    \While {$Q\ne \varnothing$}
           \State select $v\in Q$, $\Gamma_v\gets \{u~|~u\in V\setminus R, (u,v)\in S\}$
            \If {$\Gamma_v \ne \varnothing$}
                \State select $u\in \Gamma_v$, $R\gets R\cup \{u\}$, $Q\gets Q\cup \{u\}$
                \State $z_u=z_{uv}*z_v$
                \If {$z_u=+1$} \State $U\gets U\cup \{u\}$
                \Else \State $D\gets D\cup \{u\}$
                \EndIf
            \Else 
            \State $Q\gets Q\setminus \{v\}$
            \EndIf
    \EndWhile
\EndWhile
\end{algorithmic}
\end{algorithm}

 %From the construction it is obvious that the resulting graph $(V,S)$ is a forest.
 %We may call $(V,S)$ the stabilizer forest.
 The initial value of the cut weight originates from the relation between Max-Cut and the Ising model~\cite{BGJR-1988}, or simply eq.~(\ref{eq.MC-Ising}). Actually we could obtain a compact expression for the cut weight from the procedure:
 \begin{equation}
 w_c=\frac{1}{2}\sum_{1\le i< j \le n} w_{ij}+\frac{1}{2}\sum_{(i,j)\in S}|\tilde w_{ij}|.
 \end{equation}
 Here the first term denotes half of the total weight, and $\tilde w_{ij}$ indicates the weight of the edge when it is selected. Of course $\tilde w_{ij}$ may have been updated and thus differs from the original weight $w_{ij}$. From the above expression it is not difficult to see that the approximation ratio of the algorithm would be strictly larger than $1/2$. It is also not quite difficult to figure out that the choice of the DEC direction does not matter here. Thus we settle the direction problem raised in~\cite{Hassin-2021}. The time complexity of the algorithm is the same as EC and DEC. Therefore, in a naive implementation, the complexity is $\mbox{O}(n^3)$. This can be reduced to $\mbox{O}(n^2\log n)$ if we use a sorted data structure. In a more sophisticated implementation this could even be reduced to $\mbox{O}(n^2)$.

\subsubsection{Example}

We take the example in~\cite{Hassin-2021} to illustrate the procedure. We intentionally choose different contraction directions from those in DEC, to show that the results do not depend on the choices. Notice that different choices may result in different operation steps.

\begin{figure}%
\centering
\begin{minipage}{0.5\textwidth}%
\centering
\caption{Step 1}
	\includegraphics[width=0.9\textwidth]{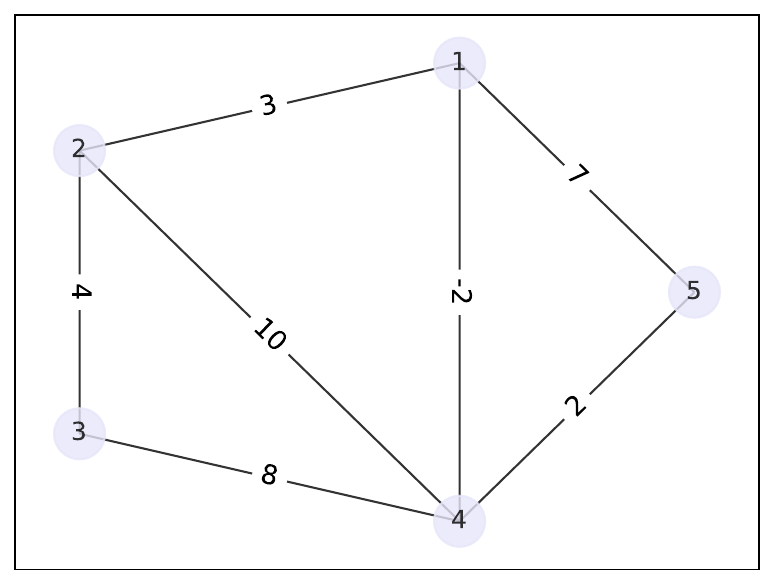}
\end{minipage}%
\begin{minipage}{0.5\textwidth}%
\centering
\caption{Step 2}
	\includegraphics[width=0.9\textwidth]{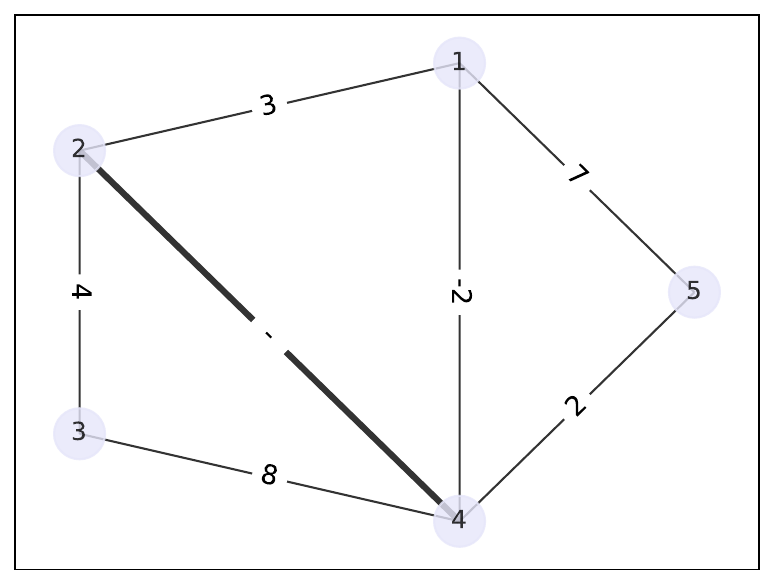}
\end{minipage}%
\\
\begin{minipage}{0.5\textwidth}%
\centering
\caption{Step 3}
	\includegraphics[width=0.9\textwidth]{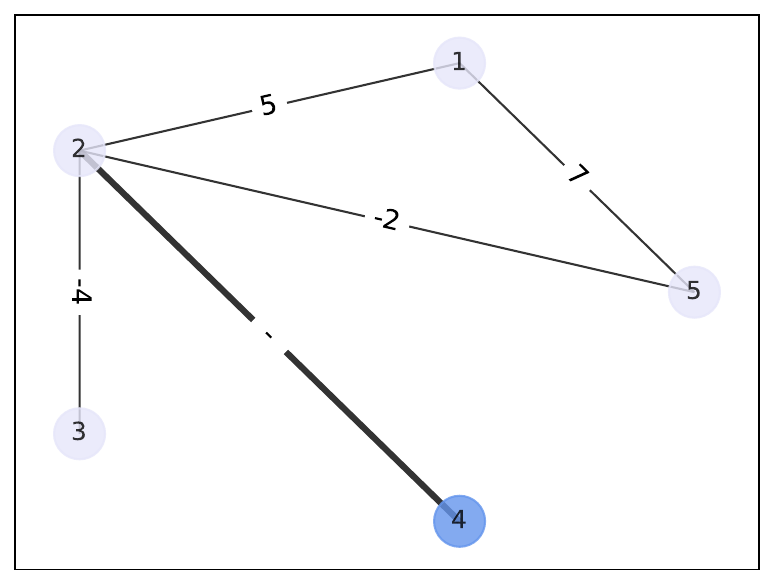}
\end{minipage}%
\begin{minipage}{0.5\textwidth}%
\centering
\caption{Step 4}
	\includegraphics[width=0.9\textwidth]{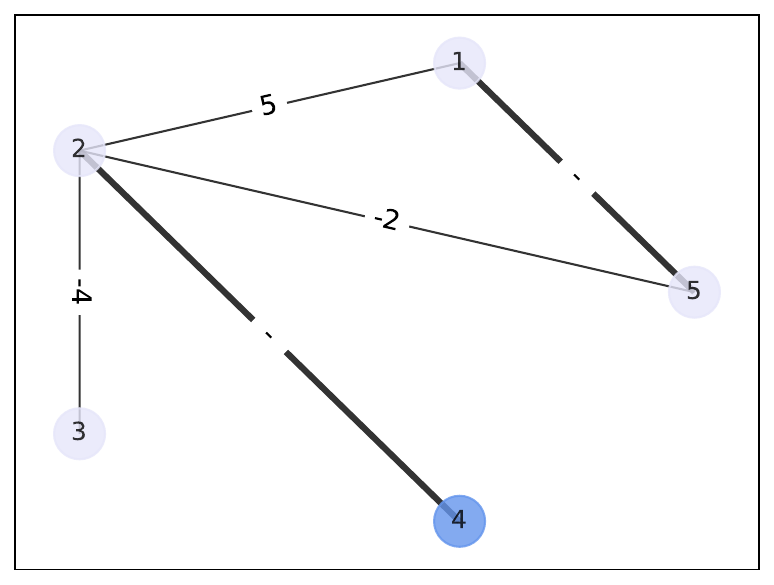}
\end{minipage}%
\\
\begin{minipage}{0.5\textwidth}%
\centering
\caption{Step 5}
	\includegraphics[width=0.9\textwidth]{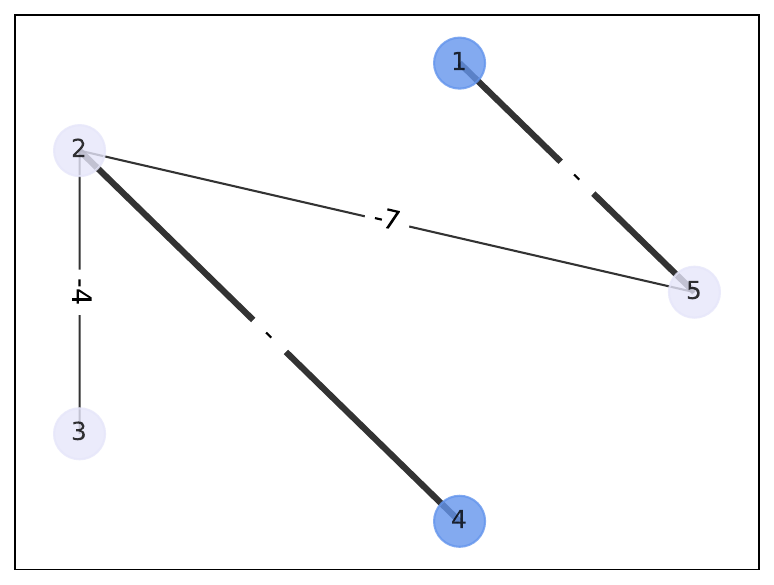}
\end{minipage}%
\begin{minipage}{0.5\textwidth}%
\centering
\caption{Step 6}
	\includegraphics[width=0.9\textwidth]{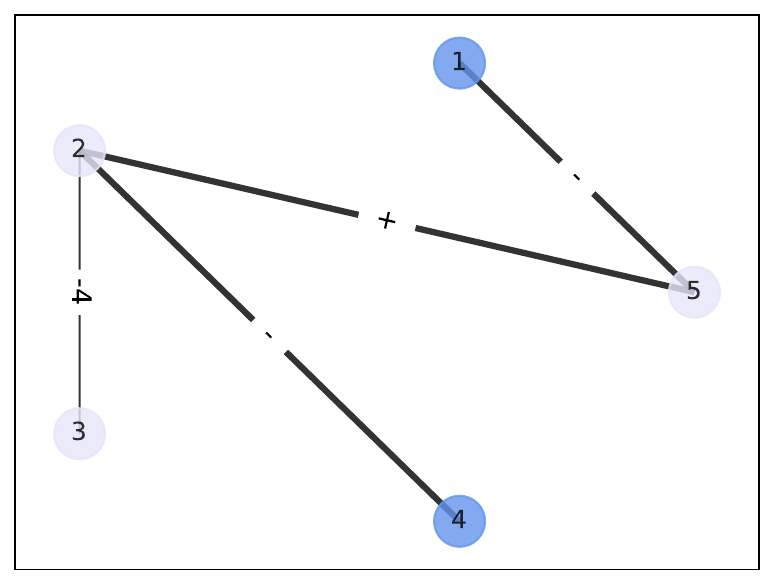}
\end{minipage}%
\\
\begin{minipage}{0.5\textwidth}%
\centering
\caption{Step 7}
	\includegraphics[width=0.9\textwidth]{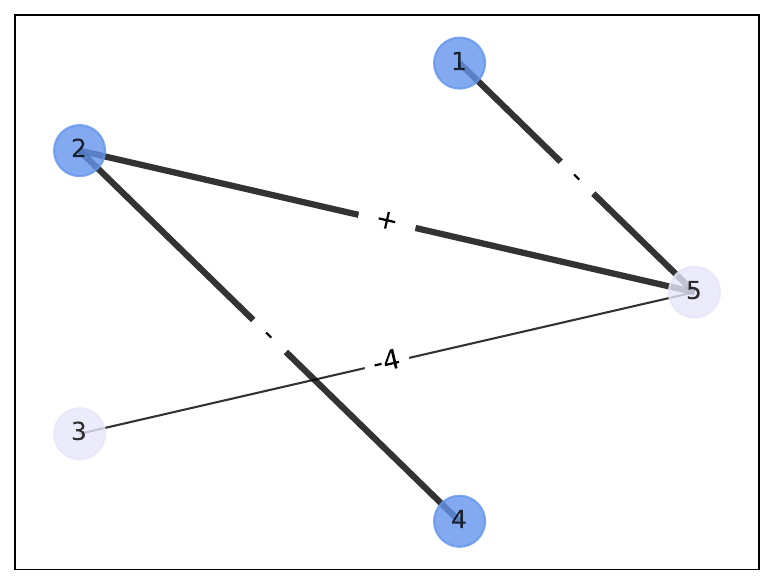}
\end{minipage}%
\begin{minipage}{0.5\textwidth}%
\centering
\caption{Step 8}
	\includegraphics[width=0.9\textwidth]{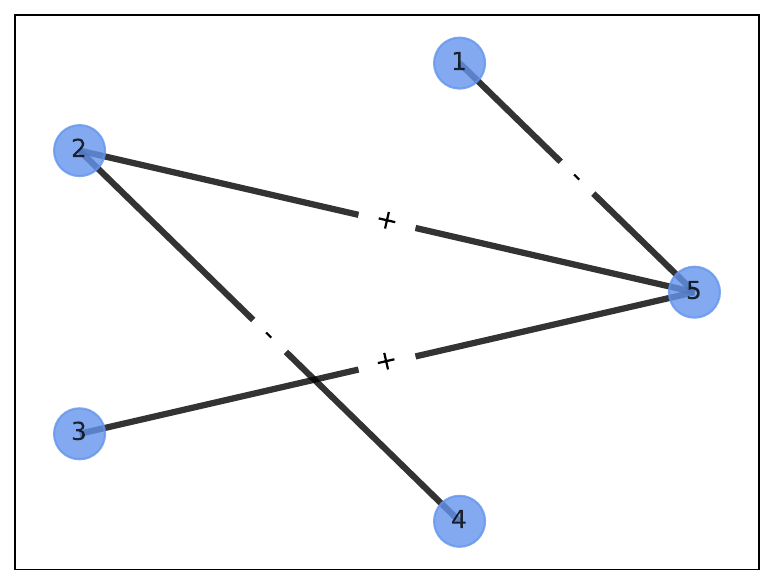}
\end{minipage}%
\caption{Illustration of the Stabilizer Heuristic procedure.}%
\label{fig:Stabilizer}%
\end{figure}

The whole procedure is listed step by step in Figure.~\ref{fig:Stabilizer}:\\
\textbf{Step 1}: input the original graph. Calculate the initial cut weight $w_c = \frac{1}{2}\sum_{1\le i< j \le n} w_{ij}=16$.\\
\textbf{Step 2}: pick out edge $(2,4)$ as the absolute value of its weight is largest. Set $z_{24}=-\frac{w_{24}}{|w_{24}|}=-1$, update the cut weight $w_c=16+|w_{24}|/2=21$.\\
\textbf{Step 3}: select vertex $4$ incident to $(2,4)$. For all the other edges incident to vertex $4$, transfer them to vertex $2$ along $(2,4)$, merge them with the original edges, and update the weights as: $w_{2k}\leftarrow w_{2k}+z_{24}w_{4k}$, $k=1,3,5$. Since $(2,5)$ does not exist in the original graph, set its initial weight to $w_{25}=0$.\\
\textbf{Step 4}: pick out $(1,5)$. Set $z_{15}=-1$, update the cut weight $w_c=+21+|w_{15}|/2=24.5$.\\
\textbf{Step 5}: select vertex $1$. Transfer $(1,2)$ to $(2,5)$, and update the weight as $w_{25}\leftarrow w_{25}-w_{12}$.\\
\textbf{Step 6}: pick out $(2,5)$, set $z_{25=}+1$. Update the cut weight $w_c=+24.5+|w_{25}|/2=28$.\\
\textbf{Step 7}: select vertex $2$. Transfer $(2,3)$ to $(3,5)$, and update the weight as $w_{35}\leftarrow w_{35}+w_{23}$. The initial weight of $(3,5)$ is again set to $w_{35}=0$, since $(3,5)$ does not exist in the original graph.\\
\textbf{Step 8}: pick out $(3,5)$, set $z_{35}=+1$. Update the cut weight $w_c=28+|w_{35}|/2=30$. \\
The resulting graph is a tree. The edge with a plus sign indicates the two vertices incident to it should be put into the same set, and the one with a minus sign indicates the vertices should be put into different sets. So we have $U=\{2,3,5\}$, and $D=\{1,4\}$. The cut weight is already calculated as $w_c=30$.

\section{Numerical Results}

Now we apply the stabilizer heuristic to some concrete instances to test its performance. The heuristic was implemented in the Python programming language (v3.7.0), and tested on a PC equipped with a 3.3GHz AMD Zen 3 Ryzen 9 processor (only use 1 core) and 2GB of RAM operating under the Microsoft Windows 11 environment. We list the computation time for some instances to show how they vary with the graph sizes.

\subsection{Comparison with SG3}

Since EC performs worse than SG3 and DEC has roughly similar performance as SG3, it will be interesting to see if the stabilizer heuristic, as a synthesis of the two, outperforms SG3. As in~\cite{Kahruman-2007}, we take some
Traveling Salesman Problem~(TSP) instances from the TSPLIB~\cite{TSPLIB} to make the comparison.

 \vskip 1em
\begin{table}[h]
\begin{center}
%{\baselineskip=18pt
\begin{tabular}{|c|c|c|c|c|c|c|}
\hline
Instance  &  Size &   Total weight  &  Stabilizer cut weight       &  Stabilizer cut ratio  &  SG3 cut ratio & Stabilizer time (ms)\\
\hline\hline
gr17           &  $17$  &  $37346$   &  $24986$ &  $0.669$  & $0.669$  &  $0.5$ \\
\hline
bayg29        &  $29$ &  $66313$  &  $42693$       &  $0.644$  &  $0.564$  &  $1.2$ \\
\hline
hk48        &  $48$ &  $1153784$  &  $771712$       &  $0.669$  &  $0.669$  &  $3.7$ \\
\hline
berlin52         &  $52$ &  $762783$  &  $470726$       &  $0.617$  &  $0.617$    &  $4.3$ \\
\hline
brazil58         &  $58$ &  $3523646$  &  $2208793$       &  $0.627$  &  $0.564$  &  $5.2$ \\
\hline
\end{tabular}
\\
\caption{Summary of results of the Stabilizer and SG3 algorithms on TSPLIB instances.}
\label{Tab.TSP0}
%}
\end{center}
\end{table}

In Table~\ref{Tab.TSP0} we list the results calculated with the stabilizer heuristic and the SG3 algorithm. We do not include the other 4 TSP instances presented in~\cite{Kahruman-2007} because the corresponding graphs have been updated. And for ``berlin52'' we round off all the weights to integers, so that the total weight agrees with that in~\cite{Kahruman-2007}. All the cut ratios are defined using the total weight as reference. Out of the 5 instances included in Table~\ref{Tab.TSP0}, the stabilizer heuristic gives the same results as SG3 for 3 of them, and much better results for the remaining 2, namely ``bayg29'' and ``brazil58''.

\subsection{Comparison with GW}

We make further comparison with the results from the Goemans-Williamson~(GW) algorithm~\cite{GW-1995} based on semi-definite programming~(SDP). We take the same TSPLIB instances as in~\cite{GW-1995} except ``gr96", which has been updated. The results from both algorithms are summarized in Table~\ref{Tab.TSP}.

\vskip 1em
\begin{table}[H]
\begin{center}
%{\baselineskip=18pt
\begin{tabular}{|c|c|c|c|c|c|}
\hline
Instance  &  Size &   Optimal cut weight  &  GW cut weight       &  Stabilizer cut weight  &  Stabilizer time (ms) \\
\hline\hline
dantzig42           &  $42$  &  $42638$   &  $42638$ &  $42638$  & $2.8$ \\
\hline
gr48         &  $48$ &  $320277$  &  $320277$       &  $320277$  &  $3.5$ \\
\hline
hk48        &  $48$ &  $771712$  &  $771712$       &  $771712$  &  $3.7$ \\
\hline
kroA100         &  $100$ &  $5897392$  &  $5897392$       &  $5897392$  &  $17.8$ \\
\hline
kroB100         &  $100$ &  $5763047$  &  $5763047$       &  $5763047$  &  $17.0$ \\
\hline
kroC100         &  $100$ &  $5890760$  &  $5890760$       &  $5890760$  &  $17.5$ \\
\hline
kroD100         &  $100$ &  $5463250$  &  $5463250$       &  $5463250$  &  $17.6$ \\
\hline
kroE100         &  $100$ &  $5986591$  &  $5986591$       &  $5986591$  &  $17.2$ \\
\hline
gr120         &  $120$ &  $2156667$  &  $2156667$       &  $2156667$  &  $25.0$ \\
\hline
\end{tabular}
\\
\caption{Summary of results of Stabilizer and GW algorithms on TSPLIB instances.}
\label{Tab.TSP}
%}
\end{center}
\end{table}

Quite unexpectedly, the stabilizer heuristic gives the optimal cut for all the instances, the same as the GW algorithm. This is a remarkable achievement, considering the performance of various other heuristics on these instances~\cite{Commander-2008}.

%One may notice that the time cost increases faster than $\mbox{O}(n^3)$. The reason is that, we have designed a general %version of the Stabilizer heuristic, which could be applied to problems beyond Max-Cut.

\subsection{Further test}

It is not difficult to devise graphs that get all the EC, DEC and Stabilizer heuristics into trouble. This is a typical example:

\begin{figure}[H]
\centering
	\includegraphics[width=0.6\textwidth]{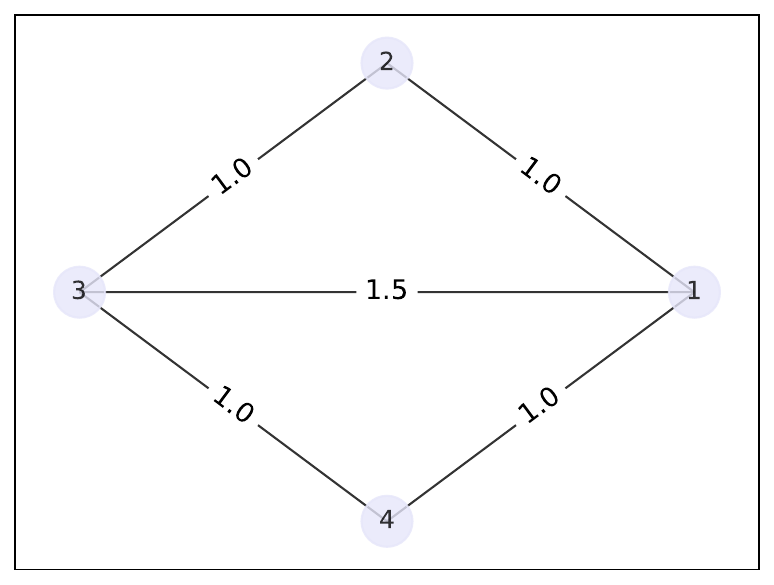}
%\vspace*{8pt}
\caption{\it A special instance for which EC, DEC and Stabilizer heuristics all fail to give the optimal cut.}\label{fig:G4-1}
\end{figure}

The optimal cut value is $4$, while all three heuristics give $3.5$. Actually, the approximation ratio can approach $3/4$ from above if the weight of the central edge approaches $1$ from above. So even if we could derive a better theoretical approximation ratio for the stabilizer heuristic, it would be at most $3/4$, much lower than the ratio $.878$ of the GW algorithm.

To see the performance of the stabilizer heuristic beyond TSPLIB instances, we test it on the Balasundaram-Butenko instances constructed in~\cite{BB-2005}. The results are summarized in Table \ref{Tab.BB}.

\begin{table}[H]
\begin{center}
%{\baselineskip=18pt
\begin{tabular}{|c|c|c|c|}
\hline
Instance  &  Size &   Optimal Value         &  Stabilizer Value  \\
\hline\hline
G5-1           &  $5$  &  $126$   &  $126$  \\
\hline
G5-2         &  $5$ &  $40$  &  $40$   \\
\hline
G8-1        &  $8$ &  $1987$  &  $1987$   \\
\hline
G8-2         &  $8$ &  $1688$  &  $1688$  \\
\hline
G10-1        &  $10$ &  $1585$  &  $1585$  \\
\hline
G10-2         &  $10$ &  $1377$  &  $1351$  \\
\hline
G15-1         &  $15$ &  $399$  &  $390$  \\
\hline
G15-2        &  $15$ &  $594$  &  $594$ \\
\hline
G20-1         &  $20$ &  $273$  &  $261$ \\
\hline
G20-2         &  $20$ &  $285$  &  $282$ \\
\hline
\end{tabular}
\\
\caption{Summary of results of Stabilizer Heuristic on Balasundaram-Butenko instances~\cite{BB-2005}.}
\label{Tab.BB}
%}
\end{center}
\end{table}

Out of the $10$ instances, the stabilizer heuristic only gives the optimal cut for $6$ instances. The maximum deviation occurs for ``G20-1'', and the deviation is about $4.4\%$. This is not too bad, compared to the results from the other approaches~\cite{Commander-2008}. One could improve these results by introducing some randomness into the algorithm. We leave this for future study.

%\newpage
\section{Discussion}

The stabilizer formalism is utilized to find approximate solutions of the Max-Cut problem. Quite unexpectedly, this leads to an elegant construction heuristic, which perfectly integrates the EC and DEC approaches together. The relation between Max-Cut and the Ising model naturally leads to an approximation ratio of at least $1/2$. For diverse instances, the new heuristic outperforms the improved versions of the SG algorithm, and can even compete with the renowned GW algorithm based on SDP. Its elegancy and remarkable performance indicates that it would be ``fast, simple and effective in practice''~\cite{Kahruman-2007}.

Why would the stabilizer formalism, a formalism developed for quantum computation specifically, finds use in such a classical problem as Max-Cut? The answer is very simple: the classical world is just a special limit of the quantum world. In principle, all classical problems could be lifted to a proper quantized version. A typical example is the quantum lifting of the Max-kSAT problem to the k-LOCAL HAMILTONIAN problem~\cite{Kitaev-2002}. Quantum thinking for the lifted problem could provide new ideas and approaches for the original problem. The stabilizer heuristic presented here is just a manifestation of such a logic.

Of course the influence should be bidirectional. Many results and proofs in~\cite{Kitaev-2002} are obtained following the classical line, especially the Cook-Levin theorem~\cite{Cook-1971,Levin-1973}. Thus one may wonder, could we develop a kind of quantum GW algorithm for Max-Cut or its quantum version?

%\section*{Framework and Resource}

%\section*{CODE AVAILABILITY}
%A preliminary version of the algorithm is developed at https://github.com/MiqroEra/Stabilizer.

%\section*{Acknowledgments}

\end{document}